\documentclass[rnote]{aa} 
\usepackage{graphicx}
\usepackage{txfonts}
\usepackage{natbib}
\newcommand{\suz}{{\it Suzaku}}
\newcommand{\cha}{{\it Chandra}}
\newcommand{\xmm}{{\it XMM-Newton}}
\newcommand{\rv}{r_{\rm vir}}
\newcommand{\om}{\Omega_{\rm m}}
\newcommand{\ol}{\Omega_{\Lambda}}
\newcommand{\hifl}{{\it HIFLUGCS}}
\newcommand{\tx}{T_{\rm X}}
\newcommand{\msun}{M_{\odot}}
\newcommand{\ro}{{\it ROSAT}}
\newcommand{\ps}{{\it PSPC}}
\begin{document}
   \title{\emph{Suzaku} measurement of Abell 2204's intracluster gas temperature
   profile out to 1800 kpc.}


   \author{T. H. Reiprich
          \inst{1}
          \and
          D. S. Hudson
          \inst{1}
          \and
          Y.-Y. Zhang
          \inst{1}
          \and
          K. Sato
          \inst{2}
          \and
          Y. Ishisaki
          \inst{3}
          \and
          A. Hoshino
          \inst{3}
          \and
          T. Ohashi
          \inst{3}
          \and
          N. Ota
          \inst{4,5}
          \and
          Y. Fujita
          \inst{6}
          }

   \offprints{T.H. Reiprich}

   \institute{Argelander Institute for Astronomy, Bonn University,
              Auf dem H\"ugel 71, D-53121 Bonn, Germany\\
              \email{thomas@reiprich.net}
         \and
              Graduate School of Natural Science and Technology,
              Kanazawa University,
              Kakuma, Kanazawa, Ishikawa, 920-1192, Japan
         \and
              Department of Physics, Tokyo Metropolitan University, 1-1 Minami-Osawa,
              Hachioji, Tokyo 192-0397, Japan
         \and
              Institute of Space and Astronautical Science (ISAS/JAXA), 3-1-1
	      Yoshinodai, Sagamihara, Kanagawa 229-8510, Japan
         \and
              Max-Planck-Institut f\"ur extraterrestrische Physik, 85748 Garching, Germany
         \and
              Department of Earth and Space Science, Graduate School of Science,
	      Osaka University, Toyonaka, Osaka 560-0043, Japan
             }

   \date{Received 2008, accepted 2009}

 
  \abstract
   {Measurements of intracluster gas temperatures out to large radii, where
   much of the galaxy cluster mass resides, are important for using
   clusters for precision cosmology and for studies of cluster physics. Previous
   attempts to measure robust temperatures at cluster virial radii have failed.}
   {The goal of this work is to measure the temperature profile of the very
   relaxed symmetric galaxy cluster Abell 2204 out to large radii, possibly
   reaching the virial radius.}
   {Taking advantage of its low particle background due to its low-Earth orbit,
   \suz\ data are used to measure the outer temperature profile of Abell 2204.
   These data are combined with \cha\ and \xmm\ data of the same cluster
   to make the connection to the inner regions, unresolved by \suz,
   and to determine the smearing due to \suz's point spread function.}
   {The temperature profile of Abell 2204 is determined from $\sim$10 kpc to
   $\sim$1800 kpc, close to an estimate of $r_{200}$ (the approximation to the
   virial radius). The temperature rises steeply from
   below 4 keV in the very center up to more than 8 keV in the intermediate
   range and then decreases again to about 4 keV at the largest radii. 
   Varying the measured particle background normalization artificially by
   $\pm$10\% does not change the results significantly.
   Several additional systematic effects are quantified, e.g., those due to
   the point spread function and astrophysical fore- and backgrounds.
   Predictions for outer temperature profiles based on hydrodynamic simulations
   show good agreement. In particular, we find the observed temperature profile
   to be slightly steeper but consistent with a drop of a factor of 0.6 from 0.3
   $r_{200}$ to $r_{200}$, as predicted by simulations.} 
   {Intracluster gas temperature measurements up to $r_{200}$ seem
   feasible with \suz, after a careful analysis of the different background
   components and the effects of the point spread function. Such
   measurements now need to be performed for a statistical sample of clusters.
   The result obtained here indicates that numerical simulations capture the
   intracluster gas physics well in cluster outskirts.}

   \keywords{X-rays: galaxies: clusters -- galaxies: clusters: individuals: Abell 2204} 

   \maketitle
%

\section{Introduction}
Cosmologically, the most important parameter of galaxy clusters is their total
gravitational mass. X-rays offer an attractive way to determine this mass
through measurements of the intracluster gas temperature and density structures.
X-rays are also a unique tool to study the physics of the hot cluster gas; e.g.,
gas temperature profiles allow constraints on (the suppression of) heat
conduction.
Consequently, constraining cluster temperature
profiles has been the subject of many (partially contradictory) works in the
recent past
\citep[e.g.,][]{mmi96,f97,mfs98,ibe99,w00,ib00,asf01,dm02,zfb04,vmm05,app05,kv05,pjk05,hr07,pbc07,smk08,lm08}.

Unfortunately, making these measurements is quite challenging in
outer cluster regions. Even with \xmm\ and \cha, it is very difficult to
determine temperature profiles reliably out to more than about 1/2 the cluster
virial radius, $\rv$ \citep[see the references above but also][for a different
view]{sas07}. As a result, only about 1/8 of the cluster volume is actually probed. The
primary reason is not an insufficient collecting area or spectral resolution of
current instruments: the limiting factor is the high particle background.
Here, the X-ray CCDs onboard \suz\ come into play. Owing to its low-Earth
orbit and short focal length, the background is much lower and more stable than
for \cha\ and \xmm\ \citep{mbi07}, making \suz\ a very promising instrument for
finally settling the cluster temperature profile debate.
And indeed, some of us have recently succeeded in measuring the temperature and metal
abundance in the outskirts of the merging clusters A399/A401 \citep{fth07}. 
Here, we go one step further and also confirm this great prospect for
outer cluster regions whose emissivity is not enhanced by merging activity and
determine the temperature profile of the regular cluster Abell 2204 out to
$\sim$1800 kpc with \suz. This radius is close to an estimate of $r_{200}$;
i.e., the radius within which the mean total density equals 200 times the
critical density, often used as approximation to the virial radius.
Note that shortly after this paper was submitted, another study
of a cluster temperature profile towards very large radii with
\suz\ was submitted by \citet{gfs08}.
 
Throughout, we assume $\om=0.3$, $\ol=0.7$, and $H_0=71$ km/s/Mpc; i.e., at the 
redshift of Abell 2204 ($z = 0.1523$), 1' = 157 kpc.

\section{Observations, Reduction, Analysis}

\subsection{\cha\ and \xmm\ data}
Abell 2204 was analyzed with \cha\ as part of the \hifl\ \citep{rb01}
follow-up. The reduction and analysis are similar to those outlined in
\citet{hrc06}. The total clean exposure of the two archival observations is 18.6
ks. Further details are described in \citet{hmr09}.

The \xmm\ analysis of Abell 2204 is published \citep{zfb08}. Extrapolating the
mass profile using the published gas density and temperature profile, we
found $r_{200}=11.75'=1840$ kpc.

\subsection{\suz\ data}
Abell 2204 was observed with \suz\ on September 17--18, 2006. We started from
the cleaned event files (processing version 1.2.2.3) and reran the ftools
xisputpixelquality and xispi, using HEAsoft 6.2 and CALDB 070409.
Using xselect, the event files were further filtered by applying the criteria
STATUS=0:65535, COR$>$6, and ELV$>$10. The good exposure amounts to about 50 ks
for each of the four XIS using both 3x3 and 5x5 editing modes. Figure~\ref{fig} 
(left) shows the combined image and the regions selected for spectral analysis.

The widths of the annuli were chosen to be about twice as large as the half power
diameter of the point spread function (PSF). Combined spectra were extracted using the 
filtered 3x3 and 5x5 editing mode events files. Using four regions from each of
the four detectors, this resulted in 16 spectra total from the source
observation. The spectra were grouped to have at least 50 counts in each energy
channel.

We checked the spectra for contamination by solar wind charge exchange emission.
The ACE SWEPAM proton flux was less than $6\times 10^{8}$ protons/s/cm$^2$
during the observation. Therefore, no strong flaring is expected \citep{fmm07}. 
Moreover, we checked the soft band ($<$2 keV) XIS-1 lightcurve of the local
background region
and did not find any indication of flaring.

Response files were created with xisrmfgen and xissimarfgen.
We generated the first Ancillary Response File (ARF) by feeding an image,
constructed from the double $\beta$ model fit to the \cha\ surface brightness
profile, into xissimarfgen. This ARF was later used to model the cluster
emission. For the second ARF, which was used when modeling the Galactic
fore- and cosmic X-ray background, we assumed a uniform photon
distribution.
We simulated 10$^7$ photons per detector using medium sampling.

Night Earth data were used to create spectra of the particle background for each
region and each detector, weighting the spectra by cutoff rigidity.\footnote{See
http://www.astro.isas.jaxa.jp/suzaku/analysis/xis/nte/ for details.
Additionally, we discarded a very small fraction of events by requiring
T\_SAA\_HXD$>$436, resulting in 795 ks total exposure time.} These particle
spectra were supplied as background spectra 
to the XSPEC (version 12.3.1) fitting routine. The cluster emission was modeled
with an absorbed thermal model (phabs*apec). The cosmic X-ray background and
Galactic emission were accounted for with an absorbed powerlaw and an unabsorbed
thermal model, respectively (phabs*pow+apec).

All 16 spectra were fitted simultaneously. We froze the hydrogen column density
at $6.07\times 10^{20}$ cm$^{-2}$ \citep{kbh05}, the cluster redshift
at $z=0.1523$ \citep{sr87}, the redshift and metal
abundance of the Galactic
component at $z=0$ and $A=1$ solar, and the powerlaw photon index at 1.41
\citep[][]{kim02}.
The remaining parameters of the fore- and background components were
left free and linked across all regions and detectors.
The temperatures, metallicities \citep{ag89}, and normalizations of the
cluster emission were left free and linked across the four detectors.
Annulus 4 served to help constrain the fore- and
background components because no significant
``contaminating'' cluster emission is expected beyond the estimated $r_{200}$,
so the normalization of the thermal cluster model was frozen at 0 in this
annulus. 

In the spectral fitting for this hot cluster, we ignored all photons with
energies $\le$0.7 keV. We did this to minimize any systematic
uncertainties due to the correction for the effective area degradation by the
contamination, due to the Galactic foreground subtraction, and due to the
influence of possible low intensity solar wind charge exchange emission.
The upper energy cut used for the fitting was set at 10 keV
(8 keV for XIS-1, due to remaining calibration uncertainties).
The Si K edge (1.8--1.9 keV) and, for XIS-0, the calibration Mn
K line (5.7--6.0 keV) were excluded. The range 7.2--8.0 keV was ignored because
of the presence of the strong instrumental Ni K line. Overall, the model
provided a good description of the data, resulting in $\chi^2_{\rm red} = 1.20$
for 4097 degrees of freedom.

\section{Results}

\begin{figure*}
\centering
\includegraphics[width=6.5cm]{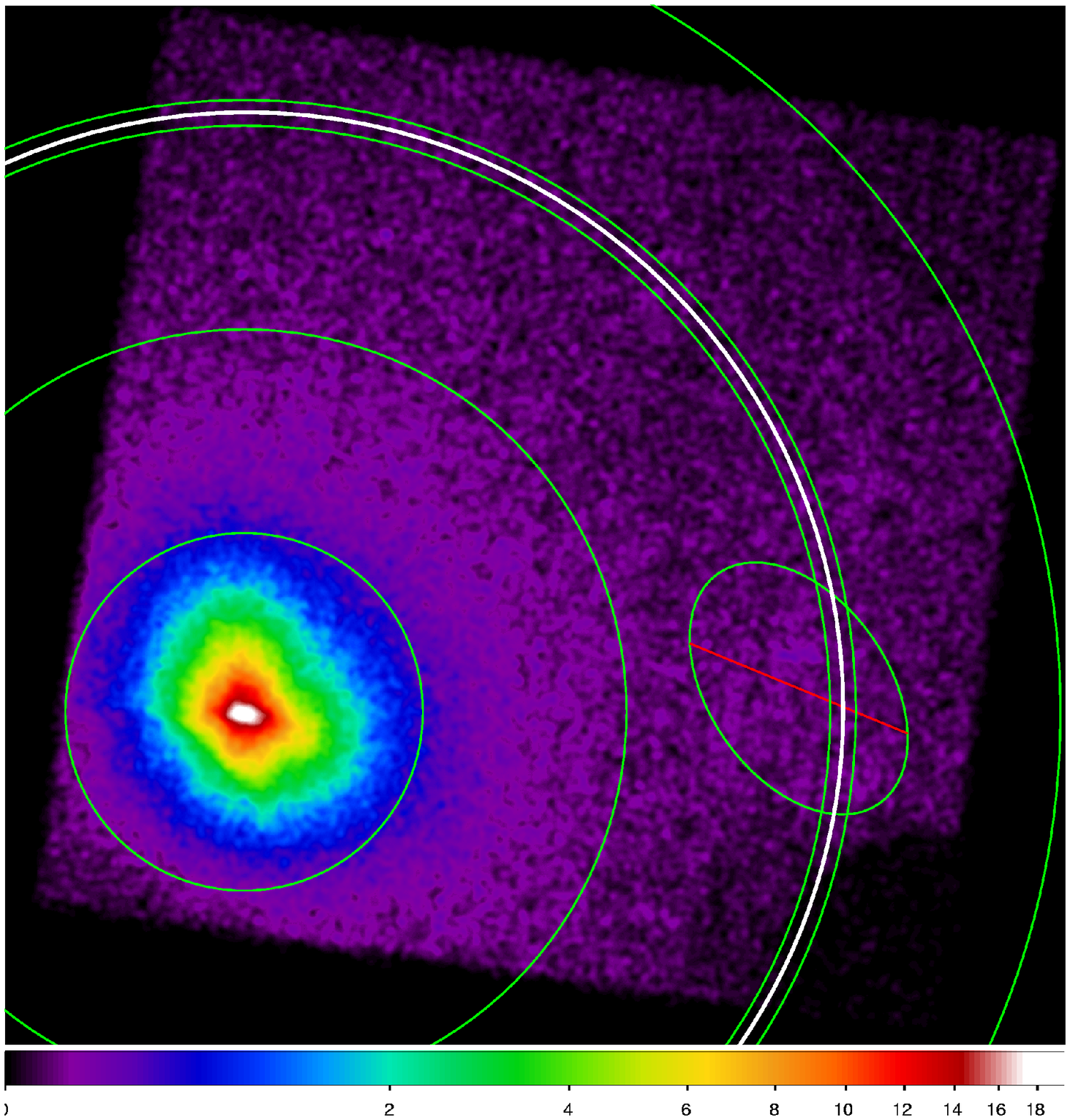}
\includegraphics[width=10.cm]{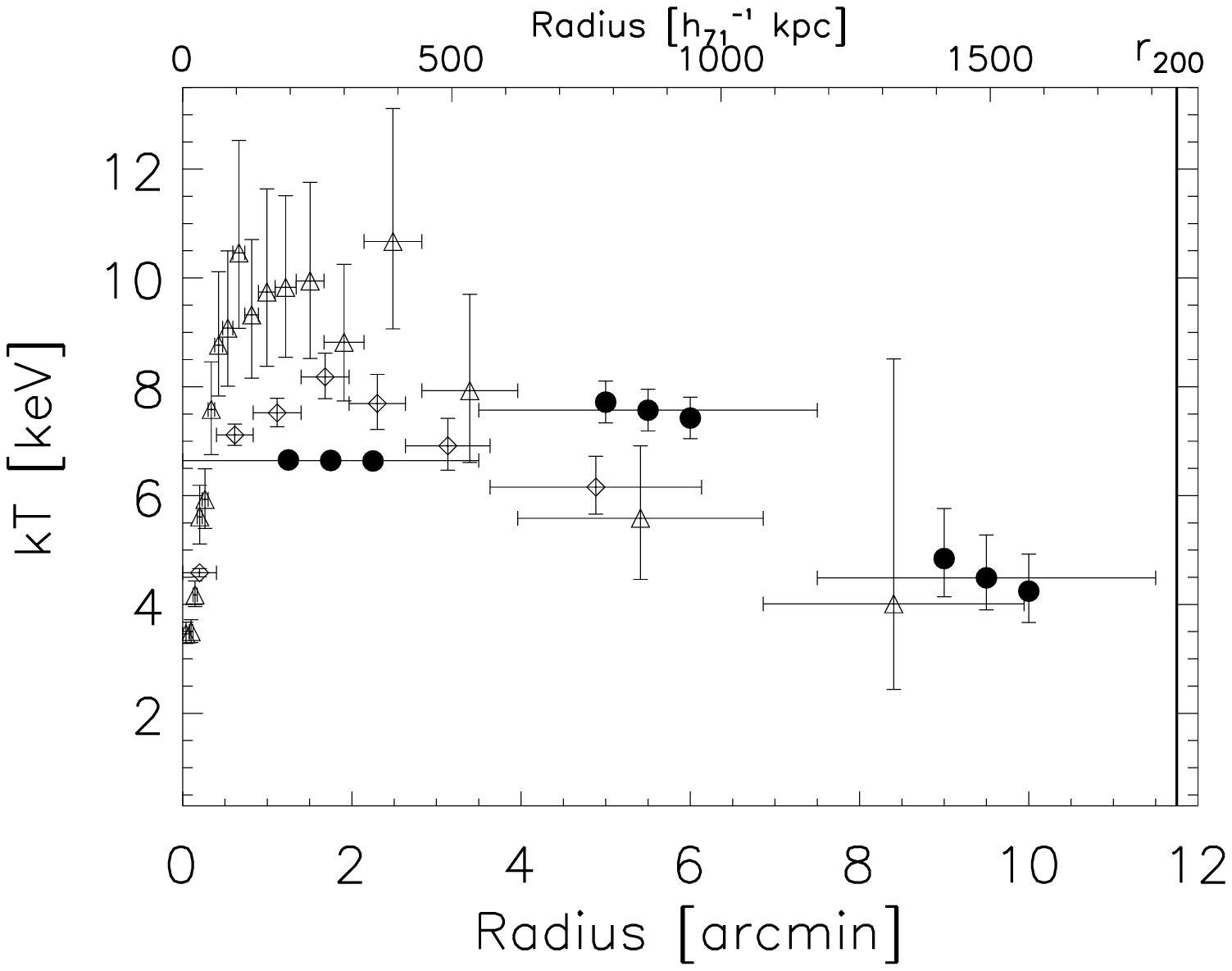}
\caption{{\it Left panel}: \suz\ image of A2204 combining all four XIS detectors. Also
indicated are the regions used for the spectral analysis (green) and the
estimated $r_{200}$ (white). The outermost annulus
(annulus 4, beyond $r_{200}$) was used to aid in the local fore- and
background estimation. The elliptical region contains emission from unrelated
sources and was excluded from the spectral analysis.
{\it Right panel}: Gas temperatures as function of radius as
measured with \suz\ (filled circles), \xmm\ (diamonds), and \cha\ (triangles).
For each \suz\ bin three temperatures are shown. The upper and lower temperature
values were obtained by artificially decreasing and increasing the particle
background normalization by 10\%, respectively. The bins are slightly offset in
this plot for clarity.
All errors are given at the 90\% confidence level.
}
\label{fig}%
\end{figure*}
%
\begin{figure*}
\centering
\includegraphics[width=7.0cm,angle=270]{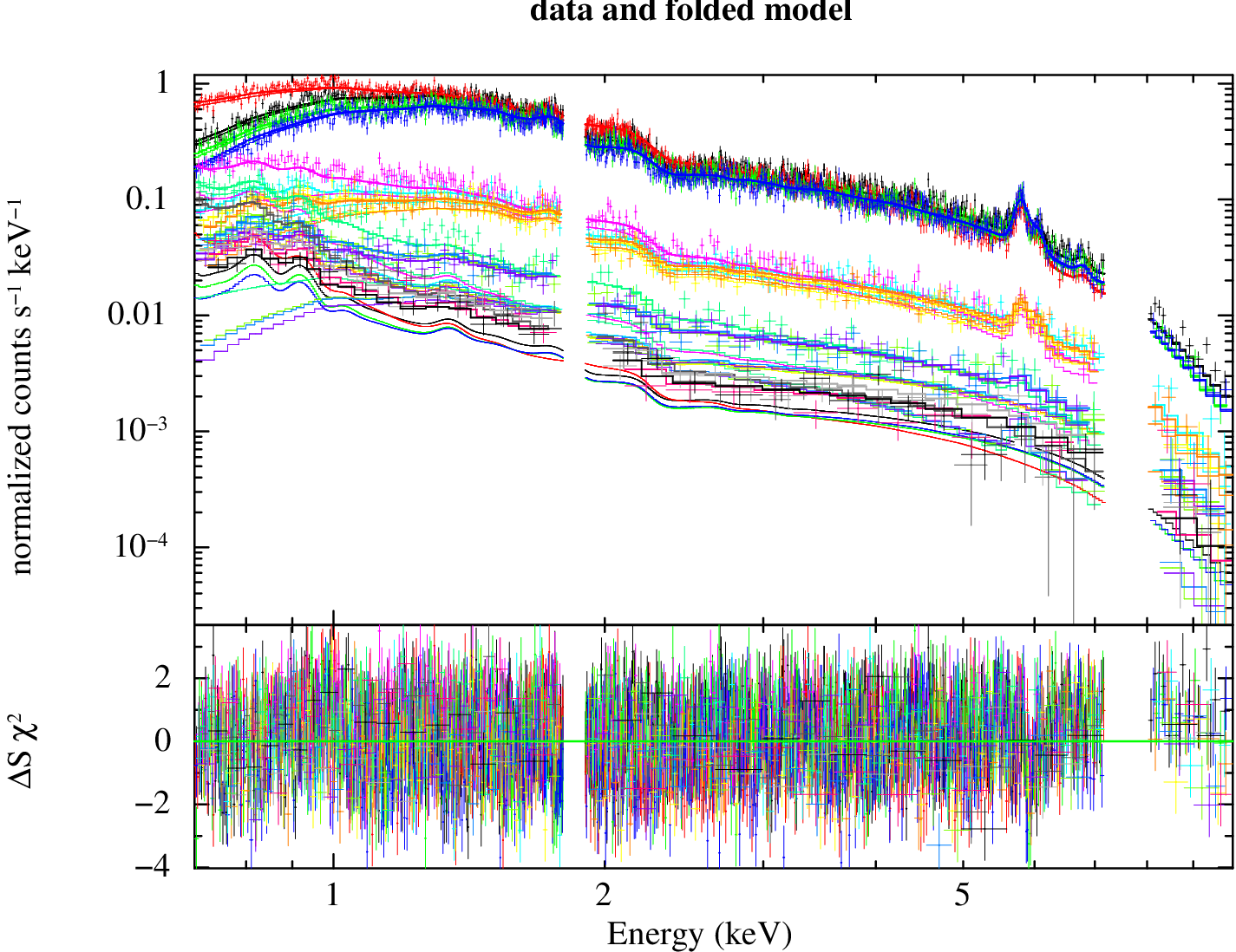}
\includegraphics[width=7.0cm,angle=270]{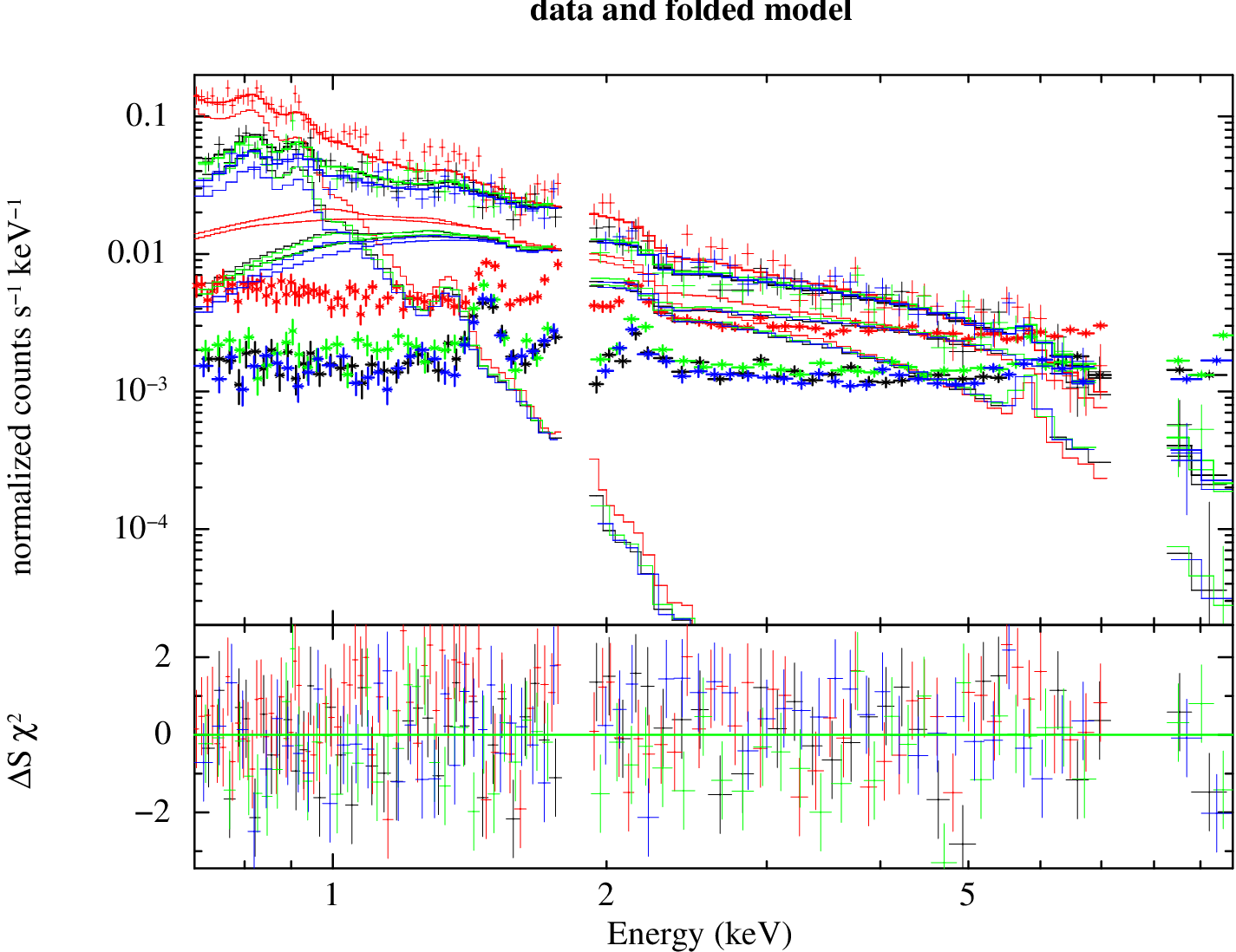}
\caption{
{\it Left panel}: Particle background subtracted \suz\ spectra
from all four regions (innermost region at the top, outermost region at the
bottom) and from all four detectors (e.g., for the innermost region: black:
XIS0, red: XIS1, green: XIS2, blue: XIS3). Also shown are 
the models (total, cluster, and combined Galactic fore- and cosmic X-ray
background) from the simultaneous fit as well as the residuals in terms of
standard deviations. There are no strong systematic deviations for any of the
regions or detectors.
{\it Right panel}: Same as left but only for bin 3 (7.5$'$--11.5$'$). In this
plot the Galactic fore- and cosmic X-ray background models are shown separately,
and the subtracted particle background is also given (by the thicker data points
at the bottom). The cluster emission dominates over the other components only in
the range $\sim$1--$1.5$ keV and even there only marginally (see text).
}
\label{spec}%
\end{figure*}

The spectra and best fit models are shown in Figure~\ref{spec}. In the left
panel, we show all 16 spectra (data points with errors) and corresponding best
fit models (solid lines). The particle background was subtracted from the data
points. Each spectrum has three model lines, one thin line representing the
cluster emission, another thin line representing the combined Galactic fore- and
cosmic X-ray background, and the thick line the sum of these (i.e., the thick
lines should match the data points).
The main purpose of this plot is to show that there are no strong residuals
anywhere.

In the right panel, we show the same plot for the most important region (bin 3).
It is instructive to compare the role of the different fore- and background
components relative to the cluster emission. Therefore, we show here also the
particle background spectra as stars with thick error bars (which were
subtracted from the data) as well as the Galactic fore- and cosmic X-ray
background components individually. For clarity, let's focus on the XIS1
spectra; i.e., the red data points and lines. The thick line matches the data
points pretty well, as it should. Now let's start at the low energies: The
uppermost thin line represents the Galactic foreground component. This component
dominates over all other components up to about 1 keV. The next line represents
the cluster component, which starts to dominate (marginally) around 1 keV until
the cosmic X-ray background becomes stronger around 1.5 keV. The particle
background starts 
to dominate over the Galactic foreground around 1.4 keV, over the cluster
component around 3 keV, and over the cosmic X-ray background between 4 and 5
keV.
This plot illustrates that a cluster temperature determination in this region
requires long exposures and is getting close to the limit of \suz's
capabilities.
In case of \cha\ or \xmm, the particle background would start to dominate
over each of the other components at much lower energies.
The same plot for bin 4 looks quite similar, there are just fewer data points
and no cluster emission.

Figure~\ref{fig} (right) shows projected gas temperature profiles measured with
\cha, \xmm, and \suz. In the innermost region ($\lesssim$0.4$'$) the \xmm\
temperature is consistent with the range of best fit temperatures obtained with
\cha\  (3.5--7.6 keV).
In the region $\sim$1--2$'$, however, \cha\ gives systematically higher
temperatures than \xmm. The primary reasons for this are, first, that the \xmm\
profile is not corrected for PSF smearing, so the temperatures in this region are
slightly underestimated; and, second, that the \cha\ effective area calibration
is, apparently, inaccurate, resulting in overestimates of the
temperature at high temperatures \citep[e.g.,][and L. David's
presentation\footnote{http://cxc.harvard.edu/ccw/proceedings/07\_proc/presentations/david/}]{smk08,svd08}.
The central \cha\ temperatures determined here are consistent with those from
\citet{sft05}.
In the region $\sim$3--6$'$ \cha\ and \xmm\ give consistent results.

The \suz\ best fit temperatures and uncertainties are given in
Tab.~\ref{tab}.
The innermost \suz\ bin covers 0$'$--3.5$'$ and, obviously, contains emission from
quite a range of temperatures. The result of the single temperature fit
($6.65_{-0.08}^{+0.08}$ keV) lies
roughly in the middle of this temperature range, as expected. The \suz\ best fit
temperature in the second bin (3.5$'$--7.5$'$, $7.57_{-0.38}^{+0.39}$ keV) is
slightly higher compared to the 
\cha\ and \xmm\  temperatures. The primary reason for this is most likely
contamination from emission from the very bright region $<$3.5$'$ due to \suz's
PSF (contamination is $\sim$50\%, see discussion below). Even though A2204
appears quite regular on large scales \citep[e.g.,][]{sbr00}, possible deviations from exact spherical symmetry of the
temperature structure may also introduce some scatter in the comparisons.

For the purpose of this paper, bin 3 covers the most important region. It
extends from 7.5$'$ to 11.5$'$; i.e., from 64\% to 98\% of the estimated $r_{200}$.
The \suz\ best fit temperature ($4.49_{-0.59}^{+0.79}$ keV) is significantly
lower than the temperatures measured further in. For \cha\ and \xmm, temperature
measurements in the low surface brightness cluster outskirts are strongly
affected by the uncertainty in the particle background. We tested the influence
of the particle background on the \suz\ results by artificially changing the
normalization of the Night Earth spectra by a very conservative
\citep[e.g.,][]{thn08} $\pm$10\% and 
repeating the fitting procedure. We found that the new best fit temperatures are
consistent with the statistical uncertainty of the standard fit
(Fig.~\ref{fig}). This demonstrates that the results obtained here are not
affected significantly by uncertainties in the particle background subtraction.

We also attempted a temperature
measurement with \cha\ in the region 6.9--9.9$'$. The \cha\ best fit temperature
is consistent with the allowed \suz\ temperature range but the
statistical uncertainty (excluding systematic effects due to the particle
background) is too large for a meaningful constraint of the temperature with
\cha.

   \begin{table}
      \caption[]{Best fit \suz\ cluster temperatures, metallicities, and the
      corresponding statistical uncertainties (90\% confidence level) of the
      four radial bins.}
     $$ 
         \begin{array}{ccccc}
            \hline
            \noalign{\smallskip}
            $Bin$ & $Radius$/['] & $Radius$/[R_{200}] & k\tx/[\mathrm{keV}] & A/[\mathrm{solar}] \\
            \noalign{\smallskip}
            \hline
            \noalign{\smallskip}
            1& $0--3.5$	& $0--0.30$& 6.65_{-0.08}^{+0.08}	& 0.43_{-0.02}^{+0.02}\\
            \noalign{\smallskip}
            2& $3.5--7.5$	& $0.30--0.64$& 7.57_{-0.38}^{+0.39}	& 0.37_{-0.07}^{+0.07}\\
            \noalign{\smallskip}
            3& $7.5--11.5$	& $0.64--0.98$ & 4.49_{-0.59}^{+0.79}	& 0.25_{-0.22}^{+0.25}\\
            \noalign{\smallskip}
            4& $12--16$	& $1.02--1.36$ & -	& -\\
            \noalign{\smallskip}
            \hline
         \end{array}
     $$ 
   \label{tab}
   \end{table}

\section{Discussion}

\begin{figure*}
\centering
\includegraphics[width=10.cm]{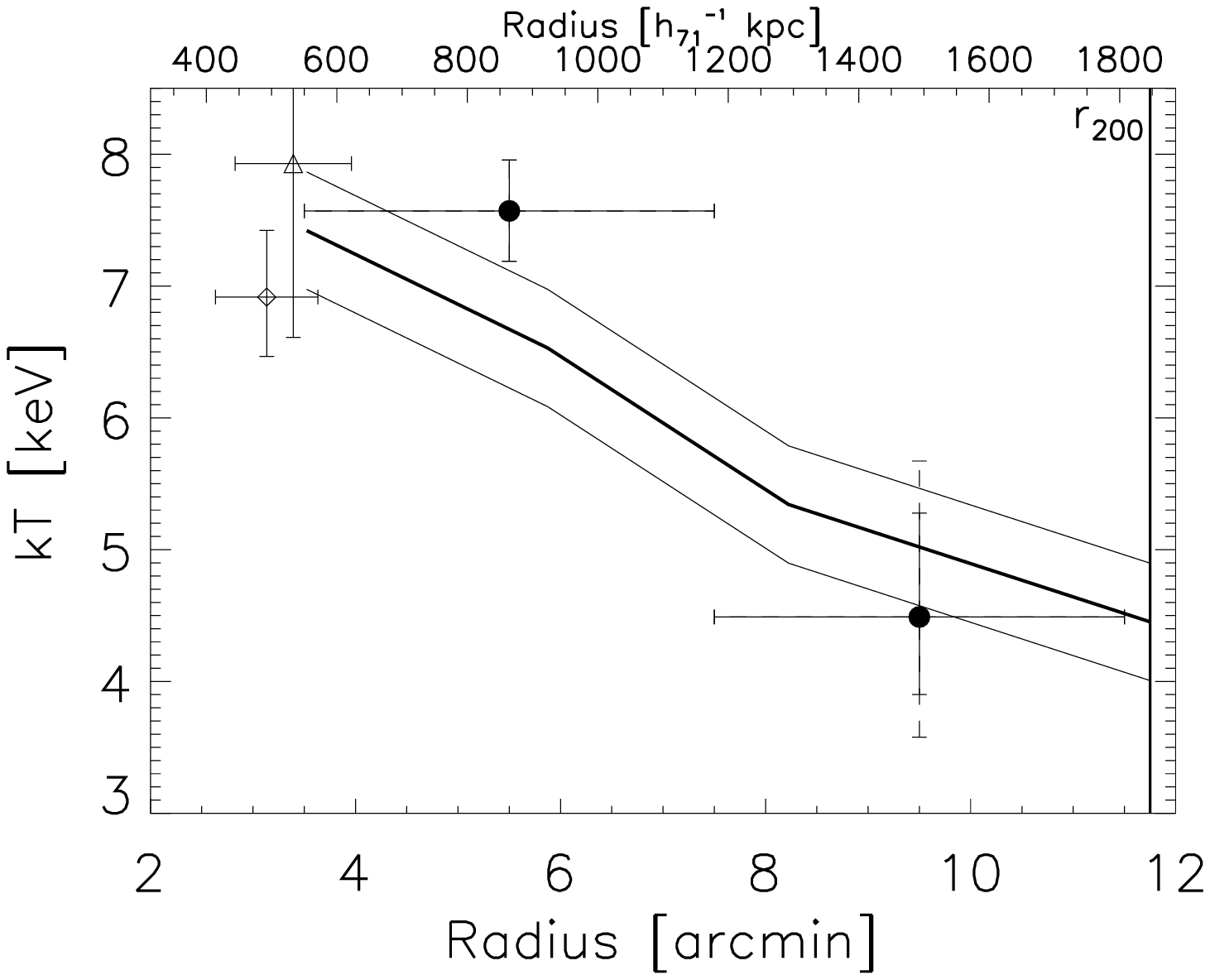}
\includegraphics[width=6.5cm,angle=270,]{0404fg3b.eps}
\caption{\emph{Left panel:}
Observed outer temperature profile compared
to profile and scatter predicted by hydrodynamical simulations of
\citet[solid lines]{red06}. Symbols have the same meaning as in
Fig.~\ref{fig}. For clarity, only the two \cha\ and \xmm\ data
points are shown that were used to determine $k\tx(0.3 r_{200})$. The other
\cha\ and \xmm\ data points further out are consistent with the
simulation results.
The dashed error bars of the outer \suz\ bin indicate the combined
statistical plus systematic error range, calculated as described in the text and
using the uncertainties given in Tab.~\ref{sys}.
\emph{Right panel:} Contamination fractions of the XIS1 bin 7.5$'$--11.5$'$
determined from ray tracing simulations. 75\% of the
photons detected in this bin also originated in this region. The energy
dependence is negligible (the lines for all five energies overlap) and the
results are very similar for the other detectors.
}
\label{fig_simu}%
\end{figure*}

We showed clear evidence that the temperature of this relaxed cluster
declines significantly when going from the inner to the outer regions.
This is expected
theoretically \citep[e.g.,][]{fwb99}. To compare, in detail, the result obtained
here to predictions, we overplotted in Fig.~\ref{fig_simu} (left) the
average temperature profile in cluster outskirts as determined with
hydrodynamical simulations of massive clusters by \citet[][Sample A in their
Tab.~2]{red06}. These authors specifically studied the regions around the virial
radius. Note that magnetic fields and cosmic rays were not included in their
simulations.
For the comparison, we used $r_{200}=11.75'$ and $k\tx(0.3 r_{200})=7.42$ keV.
The latter was calculated by taking the average of the best fit \cha\ and \xmm\
temperatures of the bins that include $0.3 r_{200}$ (shown in
Fig.~\ref{fig_simu}, left). We excluded the \suz\ measurement because its broad
PSF makes an accurate determination at this radius difficult.

The temperature of the inner \suz\ bin shown is slightly higher but consistent
with the prediction (keep in mind that, in general, regions left of bin centers
carry more weight in the temperature determination than regions right 
of bin centers, because the surface brightness decreases rapidly with radius).
As mentioned above and discussed in more detail below, this 
bin is strongly affected by PSF smearing, so the actual projected temperature
may be lower. Due to the steep surface brightness profile, deprojection would
likely result in only a minor increase of the temperature estimate.
The important outer bin is much less affected by PSF and projection effects and
is slightly lower but consistent with the prediction. In conclusion, the
observed outer temperature profile is slightly steeper but consistent with a
drop of a factor of 0.6 from 0.3 $r_{200}$ to $r_{200}$, as predicted by
simulations.

We tested illustratively what the improvement in the uncertainty of a
total mass estimate is due to the improvement of the uncertainty of the
temperature profile provided by the \suz\ data. We fitted powerlaws to
the outermost \xmm\ data point and the \cha\ and \suz\ outermost data
points, respectively, taking into account the upper and lower statistical
(90\%) errors of the latter two. Then we determined a fiducial total mass
using the best fit single $\beta$ model parameters for the density profile
\citep[from][]{rb01} and the powerlaw model parameters for the
temperature profiles, under the assumption of hydrostatic equilibrium.
Using the \cha\ temperature we found $M(<1840\,{\rm
kpc})=5.34^{+2.97}_{-2.10}\times 10^{14}\msun$. Using \suz\ we found
$M(<1840\,{\rm kpc})=6.18^{+0.64}_{-0.56}\times 10^{14}\msun$; i.e., the
uncertainty due to the temperature profile decreased from 40--60\% (\cha)\
to 10\% (\suz).
So, this represents a significant improvement of a factor of $\sim$5
(factor of $\sim$3 when taking into account the different exposure
times).
Note that a full
analysis of the total mass profile will be performed in a more in-depth
work.

In the following, we discuss and quantify several systematic effects.
We discuss these issues here at some length because this is one of the first
papers determining cluster properties out to $r_{200}$, where the surface
brightness is very low and systematic effects potentially quite important.
In summary, we found that the combination of all quantified systematic effects
is of the same order as the statistical uncertainty in the most relevant bin
(7.5$'$--11.5$'$). The individual results of the tests are summarized in
Tab.~\ref{sys}.

   \begin{table}
      \caption[]{Systematic uncertainties of the cluster temperature measurement
      in the bin 7.5$'$--11.5$'$.} 
     $$ 
         \begin{array}{p{0.7\linewidth}c}
            \hline
            \noalign{\smallskip}
            Test & \Delta k\tx/[\mathrm{keV}] \\
            \noalign{\smallskip}
            \hline
            \noalign{\smallskip}
            1 Particle background ($-$5\%)	& +0.18	\\
            \noalign{\smallskip}
            2 Particle background ($+$5\%)	& -0.17	\\
            \noalign{\smallskip}
            3 PSF (25\% contamination)	& -0.68	\\
            \noalign{\smallskip}
            4 Foreground modeling (double thermal)	& +0.61	\\
            \noalign{\smallskip}
            5 ROSAT foreground (thermal)	& +0.32	\\
            \noalign{\smallskip}
            6 ROSAT background (powerlaw)	& +0.46	\\
            \noalign{\smallskip}
            7 ARF from ROSAT surface brightness	& +0.01	\\
            \noalign{\smallskip}
            8 Extrapolated cluster emission beyond $r_{200}$	& +0.24	\\
            \noalign{\smallskip}
            \hline
         \end{array}
     $$ 
   \label{sys}
   \end{table}

We start with the particle background. The reproducibility of this background
component for \suz\ is better than about 3\% \citep[e.g.,][]{thn08}. Here, we
conservatively assumed 5\% and found best fit temperatures and errors for the
third bin as $4.66_{-0.65}^{+0.81}$ keV and $4.32_{-0.54}^{+0.77}$ keV.
So, for \suz, this effect is very small.

Next, we discuss
the influence of \suz's PSF.
We performed ray tracing simulations using the xissim tool \citep{imf07} using
$10^7$ photons per energy and detector.
We assumed Abell 2204's surface brightness profile to follow the best double
$\beta$ model fit to the \cha\ data. We followed a procedure very similar to the
one described by \citet{syi07} and 
found that the second
bin (3.5$'$--7.5$'$) is significantly (49\%) contaminated by
emission originating in the cluster center ($<$3.5$'$). The most  
important region, the third
bin (7.5$'$--11.5$'$) is contaminated by 25\%; i.e., a
relatively small fraction (see Fig.~\ref{fig_simu}, right, for the contamination
fractions of this bin determined for XIS1, as a representative example). The energy
and detector dependence of the PSF were
found to be small enough to be negligible
here. The same is true for the fine details of the model for the surface
brightness profile, since we found very similar fractions when repeating
this analysis using the best fit double $\beta$ model as obtained from \ro\ \ps\
pointed data.

Naively, one could assume that a simple PSF contamination correction could be
performed, starting with the assumption that 49\% of the emission in the
3.5$'$--7.5$'$ bin comes from a plasma at
$k\tx=6.65$ keV (Tab.~\ref{tab}).
However, this would not be accurate, as the \cha\ and \xmm\ data reveal a wide
range of temperatures at different distances from the cluster center
(Fig.~\ref{fig}). A detailed correction for the PSF effects, therefore,
requires a large number of ray tracing simulations to be performed, using fine
\cha\ temperature bins, to determine precisely how much emission from
plasma at what temperature contaminates each of the \suz\ bins.
A detailed correction for the PSF effects is beyond the scope of this paper.
Such a correction will be performed in a more in-depth analysis, using
the longer \cha\ observation of this cluster \citep{sft09} and the recently
updated effective area calibration.
Here, we follow a rather conservative route and quantify the changes in
temperature and uncertainty by assuming the temperature for the emission,
contaminating the important third bin, to lie in the range 6.0--8.0 keV.
Furthermore, we freeze the metallicities of the third bin to 0.2 and that of the
contaminating emission to 0.3. This results in the following best fit
temperatures and uncertainties: $k\tx(7.5'$--$11.5') = 4.13_{-0.77}^{+0.96}$ keV
(for $kT_{\rm contami}=6$ keV) and $k\tx(7.5'$--$11.5') = 3.81_{-0.67}^{+0.90}$ keV
(for $kT_{\rm contami}=8$ keV). Changing the metallicities of the third bin to
0.15 and 0.25 has a quite negligible influence on the best fit temperatures. The
same is true when changing the metallicity from the contaminating emission from
the second bin to 0.25 and 0.35.
In Tab.~\ref{sys} and the calculation of the total systematic uncertainty
we use the more conservative ($kT_{\rm contami}=8$ keV) result.

We specifically designed the \suz\ observation in such a way as to
have a cluster free region available to help constrain Galactic
fore- and
cosmic X-ray background directly and locally, using the same observation. This
ensures that the results are not affected by systematic calibration
differences between different satellites and varying point source
subtraction fractions. Still, we checked whether the resulting
parameter values are in a reasonable range. We found that the
normalization of the powerlaw component expected from deep CXB
studies using several different satellites is lower by factors
0.77--0.62 compared to our best fit normalization. We then froze the
powerlaw normalization to the highest \citep{vmg99} and lowest 
\citep{rgs03} normalization we found in the literature, resulting in
temperatures $6.08^{+0.98}_{-0.82}$ and $6.84^{+1.17}_{-0.72}$ keV,
respectively, in the third
bin. This procedure results in worse fits
but note that, in the latter case, the temperature is significantly
hotter than when leaving the powerlaw normalization free. Since
intensity variations of the CXB may be correlated with large scale
structure, we decided to use the independent \ro\ \ps\ observation of
Abell 2204 and fitted the same
fore-/background model in the energy band 0.3--2.4 keV
to the outskirts. We found that the normalization is fully consistent (factor
0.81--1.04) with the higher powerlaw normalization from the \suz\ data of Abell
2204. Moreover, we also found the temperature and normalization of the thermal
component to be fully consistent with the \suz\ results.
In particular, for the Galactic emission, we found temperatures of
$0.25_{-0.01}^{+0.02}$ keV (\suz)\ and $0.26_{-0.01}^{+0.01}$ keV (\ro), quite
typical of this foreground
component; i.e., there was no indication of
significant hot cluster or accretion shock emission ``contaminating'' annulus 4.
We, therefore, conclude
that our Galactic
fore- and cosmic X-ray background modeling is adequate.

Recall that the purely statistical uncertainties on these fore- and background
components are already included in our quoted statistical errors on the cluster
temperatures because in our standard analysis we performed a simultaneous fit of
all the astrophysical fore- and background components together with the cluster
component; i.e, the relevant parameters for these models were all free to vary.
For the calculation of the total systematic uncertainty, we additionally
included effects, which relate to the modeling of the Galactic thermal
emission as well as the spatial variation of the fore- and background
components, as described in the following.

Several authors used more than one spectral component to model the Galactic
foreground emission \citep[e.g.,][]{syi07,smk08,hs08}.
Therefore, we tried a two component foreground model (phabs*apec+apec) instead
of a single component model (apec) for the Suzaku observation. As best fit
temperatures for the Galactic components we found 0.28 keV for the
absorbed component and 0.13 keV for the unabsorbed one, in good
agreement with the typical values quoted by \citet{smk08}.
Furthermore, the new powerlaw model normalization (representing
the unresolved cosmic X-ray background) is only 2.5\% lower than in
the original fit. This change is much smaller than the statistical
error on this normalization.

We found that the resulting new best fit cluster temperature in the most 
important and most affected third bin (7.5$'$--11.5$'$) gets a bit higher
but not significantly so (5.10 keV). On the one hand, this lends confidence to
our approach; on the other hand, this result is also not too surprising
since in our spectral analysis we ignored all photons with energies
E$<$0.7 keV, so the cluster temperature measurements are by construction
less sensitive to uncertainties in the soft fore- and background.

Also, $\chi_{\rm red}^2$ does not change by adding this additional
model component, showing that no significant improvement can be
achieved by adding this second thermal component for the data under
consideration here.

Moreover, the relative statistical uncertainties stay very similar,
e.g., for the best fit cluster temperature in the third bin the errors
at the 90\% confidence level are $-$13\% +18\% (single thermal Galactic
foreground model) and $-$15\% +21\% (double thermal Galactic foreground
model); therefore, the simple single thermal model does not result in a
significant underestimate of statistical errors. In our systematic
error analysis we do include the model dependence in the final error.

Last not least, we tested the double thermal model in the \ro\
analysis of the foreground emission. We found very similar best fit
temperatures for the Galactic components (absorbed 0.25 keV, unabsorbed
0.13 keV), although \ro's poor energy resolution
results in significantly enlarged errors for the temperatures of the two thermal
components (the two temperature ranges overlap; degeneracies cannot be
broken, resulting in an unstable error analysis). The change in the
powerlaw normalization is again well within the statistical
uncertainty.

Next, we tested the influence of a possible spatial variation of the fore-
and background estimates. The regions selected for the fore- and
background analysis in the \ro\ observation differ from those in the
\suz\ observation. Therefore, we refit the \suz\ data but this time
not letting the powerlaw and thermal component vary freely but,
separately, freezing them to the values determined from the \ro\
observation (after correction for the different covered areas).
The best fit cluster temperatures for the third bin do not change
significantly. They are $4.81_{-0.65}^{+0.81}$ keV (foreground apec model frozen)
and $4.95_{-0.66}^{+0.85}$ keV (background powerlaw model frozen). Both effects
are included in the total systematic error analysis.

The Monte Carlo ARF calculation using xissimarfgen requires a priory knowledge
of the cluster surface brightness distribution, which we provided using a double
$\beta$ model fit to the surface brightness measured with \cha. We tested the 
influence of deviations from this assumed distribution on the cluster
temperature measurements by instead generating ARFs using the best fit double
$\beta$ model from the \ro\ observation. The resulting best fit temperature in
the third bin was almost unchanged (4.50 keV).

We assumed that the ``cluster free'' region (the forth bin) contains negligible
cluster emission and argued that the best fit temperature for the Galactic
foreground emission we found supports this assumption. 
Nevertheless, we performed an additional test to estimate the contribution of 
this assumption to the total systematic error. To this end, we determined a
rough surface brightness profile using bins 1 through 3 and extrapolated 
it by conservatively assuming the surface brightness to drop by a factor
of 5 from bin 3 to bin 4. Furthermore, we assumed a temperature of 2 keV and a
metallicity of 0.2 solar in bin 4. We then included this cluster component in
the model for bin 4, in addition to the fore- and background components, and
redid the full simultaneous fit. 
The resulting new best fit cluster temperature in bin 3 is 4.73 keV;
i.e., slightly higher than the original fit result but not
significantly so. This rather small change is expected because the
extrapolated surface brightness of this fourth bin is much lower than the
surface brightness of the powerlaw background component. Therefore, the new
normalization of the powerlaw background component is also only slightly lower
(7\%) compared to the original one.
A detailed determination and discussion of the surface brightness profile and
especially its error will be presented in a subsequent paper.
We do add the effect of the possible cluster emission in bin 4 to the
systematic error budget.

Some of the systematic effects described above raise the temperature (e.g.,
using a double thermal model for the Galactic foreground emission) others
lower it (e.g., roughly accounting for PSF effects). To estimate the total
systematic error as well as the combined statistical plus systematic error, we
used the following 
scheme. We separately added all positive and negative errors from Tab.~\ref{sys}
in quadrature, resulting in a total systematic error $_{-0.70}^{+0.88}$ keV. The
total systematic error is, therefore, slightly larger but of the same order as the
statistical uncertainty ($_{-0.59}^{+0.79}$ keV).  
Then, we added the systematic and statistical errors in quadrature, resulting
in a combined statistical plus systematic error $_{-0.91}^{+1.18}$ keV (shown as
dashed error bars in Fig.~\ref{fig_simu}).

For the comparison to simulated temperature profiles, the uncertainty of
$r_{200}$ as estimated from the extrapolated \xmm\ density and temperature
profiles may be important. For instance, using a cluster mean temperature and
the relation of \citet{emn96} would result in a larger value for $r_{200}$.
Also, \citet{app05} determined $r_{200} = 2075$ kpc using the same \xmm\ data,
probably due to the flatter temperature profile they found. 
This would make the observed temperature drop even steeper in terms of
$r_{200}$, potentially resulting in tension between observation and simulations.
We will redetermine $r_{200}$ from the cluster mass profile, taking advantage of
the information on the outer temperature from \suz\ in the more in-depth
analysis of this cluster. It is reassuring to note that further corroboration
for the $r_{200}$ estimate used here comes from the independent weak lensing
analysis of this cluster \citep{cs02}, yielding $r_{200}=11.8'$, in perfect
agreement with our estimate.

The confirmation of the predictions from hydrodynamical simulations for the gas
physics in cluster outskirts indicated here rests on the analysis of a single
cluster. What is required for a general confirmation is the analysis of a
statistical sample of clusters. In the future, this can be performed with
dedicated cluster observations
accumulating rapidly in the \suz\ archive.

\begin{acknowledgements}
THR acknowledges the hospitality of Tokyo Metropolitan University.
We thank T. Erben, G. Hasinger, O.-E. Nenestyan, and P. Schneider for help in
the early stages of this work, M. Markevitch for comments on the paper, and M.
Roncarelli for sending electronic data tables of simulated clusters.
THR, DSH, YYZ acknowledge support by the Deutsche Forschungsgemeinschaft through Emmy
Noether Research Grant RE 1462/2 and by the German BMBF through the
Verbundforschung under grant no.\ 50 OR 0601. KS acknowledges support by the
Ministry of Education, Culture, Sports, Science and Technology of Japan,
Grant-in-Aid for Scientific Research No.\ 19840043. NO thanks the Alexander von
Humboldt Foundation for support.
\end{acknowledgements}

%

\end{document}